\documentstyle[multicol,twoside,eqsecnum,epsf,aps,prb]{revtex}

\begin{document}

\title{Exact renormalization of the random transverse-field Ising spin chain\\
in the strongly ordered and strongly disordered Griffiths phases}

\author{Ferenc Igl\'oi}

\address{
Research Institute for Solid State Physics and Optics, 
H-1525 Budapest, P.O.Box 49, Hungary\\
Institute for Theoretical Physics,
Szeged University, H-6720 Szeged, Hungary\\
}

\date{\today}

\maketitle

\begin{abstract}
The real-space renormalization group (RG) treatment of random transverse-field Ising spin chains
by Fisher ({\it Phys. Rev. B{\bf 51}, 6411 (1995)}) has been extended into the strongly ordered and
strongly disordered Griffiths phases and asymptotically exact results are obtained.
In the non-critical region the asymmetry of the renormalization of the couplings and the
transverse fields is related to a non-linear quantum control parameter, $\Delta$, which is
a natural measure of the distance from the quantum critical point. $\Delta$, which is found to
stay invariant along the RG trajectories and has been expressed by the initial disorder
distributions, stands in the singularity exponents of different physical quantities
(magnetization, susceptibility, specific heat, etc), which are exactly calculated. In this
way we have observed a weak-universality scenario: the Griffiths-McCoy singularities
does not depend on the form of the disorder, provided the non-linear quantum
control parameter has the same value.
The exact scaling function of the magnetization with a small applied magnetic
field is calculated and the critical point magnetization singularity is
determined in a simple, direct way.
\end{abstract}

\pacs{75.50.Lk, 05.50.+q, 64.60.Ak, 68.35.Rh} 

\newcommand{\bc}{\begin{center}}
\newcommand{\ec}{\end{center}}
\newcommand{\be}{\begin{equation}}
\newcommand{\ee}{\end{equation}}
\newcommand{\ba}{\begin{array}}
\newcommand{\ea}{\end{array}}
\newcommand{\beqn}{\begin{eqnarray}}
\newcommand{\eeqn}{\end{eqnarray}}
\begin{multicols}{2}
\narrowtext

\section{Introduction}

In statistical physics and in the theory of interacting many body systems exact solutions
are of great importance, especially in the vicinity of singular points,
such as at phase transitions. They provide physical insight about the cooperative processes and their
results could be used as testing ground for different approximations and numerical methods.
Exact solutions about models with short range interactions and in the presence of quenched
randomness are scarce which has greatly hampered our understanding about collective
phenomena in disordered systems. At present remarkable examples about exactly soluble problems
in the above class are the
critical behavior of low-dimensional random quantum systems\cite{bhatt}. Here the interplay of
quenched disorder, quantum fluctuations and correlations can be systematically studied
within a real-space renormalization group (RG) scheme, which is expected to lead to
asymptotically exact results, at least for strong enough disorder.
The prototype of such type of random quantum systems is the random transverse-field Ising
spin chain (RTIC)
for which perhaps the most detailed analytical and numerical information is available,
as far as the random quantum critical behavior is concerned. The RTIC is
defined by the Hamiltonian:
\be
\hat{H}=-\sum_l J_l \sigma_l^x \sigma_{l+1}^x-\sum_l h_l \sigma_l^z - H \sum_l \sigma_l^x\;,
\label{hamilton}
\ee
in terms of the $\sigma_l^{x,z}$ Pauli matrices at site, $l$, whereas the
transverse fields, $h_l>0$, and the couplings, $J_l>0$,
are independent random variables taken from the (initial) distributions,
$P_{\rm in}(h){\rm d} h$ and $R_{\rm in}(J){\rm d} J$, respectively. The (linear)
quantum control parameter of the model is defined as\cite{pfeuty}
\be
\delta={[\ln h]_{\rm av}-[\ln J]_{\rm av} \over \rm{var}[\ln h]+\rm{var}[\ln J]}\;.
\label{delta}
\ee
where ${\rm var}(x)$ is the variance of $x$ and here and
in the following we use $[ \dots ]_{\rm av}$ to denote averaging over quenched disorder.
The quantum critical point at $\delta=0$ separates the ferromagnetic ($\delta<0$) and
the paramagnetic ($\delta>0$) phases.

Some previously known exact results about the surface magnetization\cite{mccoy,mccoywu} and
about typical correlations\cite{shankar} have been greatly extended by
Fisher\cite{fisher} using the strong disorder RG method, originally introduced by
Ma, Dasgupta and Hu\cite{mdh}.
Fisher has shown
that at the critical point the distribution function of the couplings and that of the
transverse fields broaden without limits as the energy scale, $\Omega$, as defined by the
strongest bond or transverse field, is lowered. Therefore, as the fixed point of the
transformation with $\Omega=0$ is approached
the disorder becomes stronger and stronger, so that in this, so called, infinite randomness
fixed point (IRFP)
the ratio of typical couplings and transverse fields at neighboring sites
is either zero or tends to infinity. As a consequence in the IRFP the RG transformation becomes
asymptotically exact and the fixed-point RG equations for the RTIC can be solved, in large
extent, analytically. From the RG treatment and from other
analytical results\cite{bigpaper} we have a clear physical picture about the origin
of the critical behavior of the RTIC, which is most probably also relevant for other low-dimensional
random quantum critical systems, having an IRFP. The critical properties
of these systems are determined by the so called {\it rare events},
which are realizations occouring with a vanishing probability, but which dominate the {\it average}
properties. In contrary, the {\it typical realizations}, which appear with probability one
have a vanishing contribution to the average critical quantities. The mathematical origin of
the solubility of these models is connected with the above observation, since it is enough only to
deal with the rare events and an overwhelming part of the realizations are irrelevant
in respect of the average critical properties.

In the following we list the existing results about the singular behavior of the RTIC. 
At $T=0$ and without an external field, $H=0$, the average magnetization of the surface spin,
$m_s$, asymptotically vanishes as $m_s \sim (-\delta)^{\beta_s}$,
with an exponent\cite{mccoy,mccoywu}:
\be
\beta_s=1\;,
\label{betas}
\ee
whereas the same behavior for the bulk magnetization, $m$, involves the bulk exponent\cite{fisher},
$\beta$:
\be
\beta=2-\Phi,\quad \Phi=\frac{1+\sqrt{5}}{2}\;.
\label{beta}
\ee
Average correlations, $G(r)=[\langle \sigma_l^x \sigma_{l+r}^x \rangle]_{\rm av}$, outside
the critical point decay exponentially and the correlation length, $\xi$, asymptotically
diverges in the vicinity of the critical point as\cite{fisher}
\be
\xi \sim |\delta|^{-\nu},\quad \nu=2\;.
\label{nu}
\ee
On the other hand the {\it typical} correlation length, $\xi_{typ}$, defined through the relation
$[\ln(\langle \sigma_l^x \sigma_{l+r}^x \rangle)]_{\rm av} \sim -r/\xi_{typ}$ involves another
exponent\cite{shankar}, $\nu_{typ}=1$. At the critical point average correlations decay as a
power, $G(r)\sim r^{-2x}$, and the scaling dimension of the bulk magnetization, $x$, satisfies
the scaling relation $x=\beta/\nu$. Similarly, for end-to-end critical correlations the
corresponding scaling dimension, $x_s$, is expressed by $x_s=\beta_s/\nu$.

In a quantum system statical and dynamical correlations are inherently related. In the RTIC
at the critical point dynamical scaling is strongly anisotropic, the relevant (imaginary)
time scale the relaxation time, $t_r$, is related by the length-scale, $\xi$, as
\be
\ln t_r \sim \xi^{1/2}\;.
\label{scales}
\ee

In the RG study Fisher has also considered the {\it weakly ordered and weakly disordered}
Griffiths phases\cite{griffiths}, which are situated in the vicinity of the critical point.
He found an anisotropic
scaling relation, $t_r \sim \xi^{1/2|\delta|}$, which can be interpreted as a $\delta$ dependent
dynamical exponent, z, which in leading order is given by
\be
z\approx \frac{1}{2 |\delta|},\quad |\delta| \ll 1\;.
\label{z_appr}
\ee
In the presence of a small external field, $H \ll 1$, and in the vicinity of the critical
point, $|\delta| \ll 1$, but with a finite combination of, $\gamma=\delta \ln(H)=O(1)$,
Fisher has obtained the exact scaling function of the magnetization of the form:
\be
m(\delta,H) \approx m_0 \left[\ln(H_0/H)\right]^{\Phi-2} \tilde{m}[\delta \ln(H_0/H)]\;,
\label{m_scaling_f}
\ee
where $m_0$ and $H_0$ are dimensional constants, and $\tilde{m}(\gamma)$,
is given in Ref.\onlinecite{fisher}.

In the present paper we extend the RG treatment by Fisher into the entire Griffiths
region. which for some type of initial disorder distribution could cover the entire
off-critical region, $ 0 < |\delta| < \infty$. Our analytical results in the Griffiths phases
are asymptotically exact in the same sense as argued by Fisher at the critical point.
Here we summarize our main findings. A short account of our results, especially about
renormalization of couplings and transverse fields, has been announced in Ref.\onlinecite{ijl01}.

The asymmetry in the renormalization of the couplings and the transverse fields is
related to a {\it non-linear} quantum control parameter,
$\Delta$, which is defined as the root of the following equation:
\be
\left[\left(\frac{J^2}{h^2} \right)^{\Delta} \right]_{\rm av}=1\;.
\label{Delta}
\ee
We have shown that $\Delta$ stays invariant along the RG trajectories (i.e. as the
energy scale, $\Omega$, is lowered) and so can be expressed by 
initial disorder distributions. Note that for very different disorder distributions
one might have the same $\Delta$, provided the distributions have the same
form of asymmetry.
The dynamical exponent, $z$, is simply expressed by the non-linear quantum control parameter as:
\be
z= \frac{1}{2 |\Delta|}\;,
\label{z_exact}
\ee
which is an exact relation in the entire Griffiths region. We have calculated the singular behavior
of different physical quantities (magnetization, susceptibility, specific heat, etc) and the
singularity exponents are all expressed by the non-linear quantum control parameter, $\Delta$.
In this way we have demonstrated a weak-universality scenario: details
of the disorder distributions are irrelevant for the Griffiths-McCoy singularities, provided
the non-linear quantum control parameter has the same value.

In the vicinity of the random quantum critical point the non-linear and the linear quantum
control parameters are asymptotically identical,
\be
\Delta=\delta+O(\delta^2)\;,
\label{Delta_delta}
\ee
consequently in the linear $\delta$ limit
from our formulae we can recover Fisher's results\cite{fisher} about the {\it weakly} disordered
and {\it weakly} ordered Griffiths phases. From our results one can also obtain the scaling function
of the magnetization as a function of a
small applied magnetic field, $H$, which in the paramagnetic phase is given in the functional form:
\be
m(\Delta,H)=m_D \left(\frac{H}{H_D}\right)^{2 \Delta} \tilde{m}\left[\frac{\Delta}{\Delta_D},
\left(\frac{H}{H_D}\right)^{2 \Delta}\right]\;,
\label{m_scaling}
\ee
where $m_D$, $H_D$ and $\Delta_D$ are non-universal dimensional constants. Here $\Delta_D$
is a non-universal parameter, which depends on the details of the distribution of the
disorder. At the critical point, $\Delta=0$, our result in Eq.(\ref{m_scaling}) goes over
Fisher's result in Eq.(\ref{m_scaling_f}).

The structure of the paper is the following. The RG equations and their solution for the fixed
point distribution of the couplings and the transverse fields are presented in Sec. 2. Renormalization
of lengths and magnetic moments are given in Sec. 3. The scaling behavior of different
thermodynamic quantities in the presence of a finite external magnetic field or
at small, but non-zero temperature is calculated in Sec. 4. We conclude our paper with a discussion
about possible extension of our results to other problems in Sec. 5. Some detailed
calculations about the distribution function of lengths are given in the Appendix.

\section{Renormalization of couplings and transverse fields}

Here we consider the RTIC as given by the Hamiltonian in Eq.(\ref{hamilton}) without
external magnetic field, i.e. $H=0$. To a spin at lattice site, $l$, we assign a magnetic
moment, $\mu_l$, and a length, $l^s_l$, while the transverse field acting on this spin is
denoted by $h_l$. Similarly, to the $l$-th bond, connecting lattice sites $l$ and $l+1$, we assign
the length, $l^b_l$, and the associated coupling is denoted by $J_l$. In the initial situation
$l^s_l=l^b_l=1/2$, $\mu_l=1$ and the couplings and fields are taken from the initial
distributions, $P_{\rm in}(h){\rm d} h$ and $R_{\rm in}(J){\rm d} J$, respectively.

During renormalization the strongest term in the Hamiltonian, coupling or transverse-field,
of strength $\Omega$ is successively decimated out and the neighboring transverse-fields or
couplings are replaced by weaker ones, which are generated by a second order perturbation
calculation. If the strongest term is a coupling, say $\Omega=J_l$, then the two spins connected
by $J_l$ flip coherently in a longitudinal field, thus they act as an effective, composite spin
having the renormalized parameters:
\be
\tilde{h}= \frac{h_l h_{l+1}}{J_l},\quad \tilde{l}^b=l_l^s+l_l^b+l_{l+1}^s,
\quad \tilde{\mu}=\mu_l+\mu_{l+1} \;.
\label{hdecimation}
\ee
On the other hand, if the strongest term in the Hamiltonian is a transverse-field,
say $\Omega=h_l$, then the state of this spin in a small longitudinal field is practically
unchanged, thus its contribution to the susceptibility is negligible. Consequently in
magnetic point of view this spin can be decimated out, and the renormalized parameters of
the new effective bond connecting sites $l-1$ and $l+1$ are given by:
\be
\tilde{J}= \frac{J_{l-1} J_{l}}{h_l},\quad \tilde{l}^s=l_{l-1}^b+l_l^s+l_{l}^b\;.
\label{Jdecimation}
\ee
Note that the decimation equations for $\tilde{h}$ and $\tilde{J}$ are related through
duality.

During renormalization the energy scale is reduced and the joint distribution functions,
such as for the spins, $P(h,l^s,\mu;\Omega){\rm d}h {\rm d} l^s {\rm d} \mu$,
and that for the bonds, $R_l(J,l^b;\Omega){\rm d} J {\rm d} l^b $, are also $\Omega$ dependent.
Generally we deal with the following reduced distribution functions:
\beqn
P_0(h,\Omega)=\int \int P(h,l^s,\mu;\Omega) {\rm d} l^s {\rm d} \mu
\nonumber\\
P_l(h,l^s,\Omega)=\int P(h,l^s,\mu;\Omega) {\rm d} \mu
\nonumber\\
P_{\mu}(h,\mu,\Omega)=\int P(h,l^s,\mu;\Omega) {\rm d} l^s
\nonumber\\
R_{0}(J,\Omega)=\int R_l(J,l^b;\Omega) {\rm d} l^b\;,
\label{distributions}
\eeqn
all of which are normalized. In this Section we consider the distribution of transverse fields
and couplings, so that we work with $P_0(h,\Omega)$ and $R_{0}(J,\Omega)$, whereas the other
joint distributions, which are connected to the size of average lengths and average moments will be
considered in the following Section.

We start to calculate the variation of the distribution function of transverse fields, 
${\rm d}P_0(h,\Omega)$, when the energy scale is lowered by $\Omega \to \Omega - {\rm d}\Omega$,
which amounts to eliminate a fraction of ${\rm d} \Omega [ P_0(\Omega,\Omega)+R_0(\Omega,\Omega)]$
spins. Here one should take into account the fact that as a strong bond is decimated out
two original fields are also eliminated
and one new is created, the strength of which is given in Eq.(\ref{hdecimation}). Since the new
distribution function should also be normalized we arrive to the equation:
\begin{eqnarray}
\frac{{\rm d} P_0}{{\rm d} \Omega}=P_0(h,\Omega)\left[R_0(\Omega,\Omega)-P_0(\Omega,\Omega)\right]
\nonumber\\
-R_0(\Omega,\Omega) \int_{h }^\Omega {\rm d} h' P_0(h',\Omega)
P_0(\frac{h}{h'}\Omega ,\Omega) \frac{\Omega}{h'}
\;.
\label{Pdiff}
\end{eqnarray}
One can similarly derive the evaluation equation of the coupling distribution:
\begin{eqnarray}
\frac{{\rm d} R_0}{{\rm d} \Omega}=R_0(J,\Omega)\left[P_0(\Omega,\Omega)-R_0(\Omega,\Omega)\right]
\nonumber\\
-P_0(\Omega,\Omega) \int_{J }^\Omega {\rm d} J' R_0(J',\Omega)
R_0(\frac{J}{J'}\Omega ,\Omega) \frac{\Omega}{J'}
\;,
\label{Rdiff}
\end{eqnarray}
which follows simply from Eq.(\ref{Pdiff}) by duality, which amounts to interchange $h \leftrightarrow J$
and $P_0 \leftrightarrow R_0$. The two integro-differential equations in Eqs.(\ref{Pdiff}) and (\ref{Rdiff})
have to be supplemented by the initial conditions, represented by the distributions $P_{\rm in}(h)$ and
$R_{\rm in}(J)$.

\subsection{Fixed-point solution}

A special solution to the problem in Eqs.(\ref{Pdiff}) and (\ref{Rdiff}) is given by the functions:
\begin{eqnarray}
P_0(h,\Omega)&=&\frac{p_0(\Omega)}{\Omega}\left(\frac{\Omega}{h}\right)^{1-p_0(\Omega)}
\label{Psol}\\
R_0(J,\Omega)&=&\frac{r_0(\Omega)}{\Omega}\left(\frac{\Omega}{J}\right)^{1-r_0(\Omega)}
\label{Rsol}
\;,
\end{eqnarray}
thus they depend only on the values of the distributions at their edges, at
$P_0(\Omega,\Omega)=p_0/\Omega$ and at $R_0(\Omega,\Omega)=r_0/\Omega$.
At the end of the Section we present arguments, that
this special solution represents the true solution of the problem at the
fixed point, i.e. as $\Omega \to 0$. Later we also show how the parameters
of the special solution can be related with the initial distributions,
$P_{\rm in}(h)$ and $R_{\rm in}(J)$.

Putting Eqs.(\ref{Psol}) and (\ref{Rsol}) into Eq.(\ref{Pdiff})
we obtain:
\begin{equation}
\left[p_0 r_0 - \Omega \frac{{\rm d} p_0}{{\rm d} \Omega}
\right]\left[\ln \frac{\Omega}{h}-\frac{1}{p_0} \right]=0
\;,
\label{Rint1}
\end{equation}
which leads to the ordinary differential equation:
\be
\frac{{\rm d} p_0}{{\rm d}\Gamma}=-p_0 r_0\;,
\label{Pedge}
\end{equation}
in terms of the log-energy variable, $\Gamma=-\ln \Omega$. Similarly, from
Eq.(\ref{Rdiff}) we obtain for the
edge-parameter, $r_0$:
\be
\frac{{\rm d} r_0}{{\rm d}\Gamma}=-r_0 p_0\;.
\label{Redge}
\end{equation}
Subtracting Eq.(\ref{Pedge}) from Eq.(\ref{Redge}) we obtain that $p_0$ and $r_0$ differ from each
other by a constant, $2 \Delta$,
\be
p_0-r_0=2 \Delta\;.
\label{diff}
\ee
Thus in terms of the variable
\be
y_0=p_0-\Delta=r_0+\Delta\;
\label{y_0}
\ee
we obtain one differential equation:
\begin{equation}
\frac{{\rm d} y_0}{{\rm d} \Gamma} + y_0^2=\Delta^2\;.
\label{dif3}
\end{equation}
Here we note that $\Delta$ is related to the asymmetry in the renormalization of couplings
and transverse fields and its value, which can be expressed by the initial
distributions, will be determined later. At the critical point, where the distributions
of the transverse fields
and that of the couplings evolve to the same limiting function as $\Omega \to 0$, we have
$\Delta=0$. In the paramagnetic phase,
where according to Eq.(\ref{delta}), transverse fields
in average are stronger than the average couplings, $\Delta>0$, whereas in the ferromagnetic
phase we have the opposite situation, $\Delta<0$.

{\it At the critical point}, $\Delta=0$, the solution to Eq.(\ref{dif3}) is given by:
\begin{equation}
y_0=p_0=r_0=\frac{1}{\Gamma-\Gamma_0}=\frac{1}{\ln(\Omega_0/\Omega)},
\quad \delta=\Delta=0\;,
\label{sol0}
\end{equation}
where $\Gamma_0=-\ln \Omega_0$ is a reference (log)energy scale. It is instructive
to consider the distribution of the reduced log-coupling variable $\eta=-(\ln \Omega 
-\ln h)/\ln \Omega=-(\ln \Omega  -\ln J)/\ln \Omega$, which is given from
Eqs.(\ref{Psol}) and (\ref{sol0}) as
\begin{equation}
\rho(\eta){\rm d} \eta = \exp(-\eta){\rm d} \eta \;.
\label{soleta}
\end{equation}
This is just the critical point solution of the RTIC by Fisher\cite{fisher}.

The solution to Eq.(\ref{dif3}) {\it in the off-critical region}, $\Delta \ne 0$, is
given by:
\beqn
y_0&=&\frac{\Delta \overline{y}_0+\Delta^2 {\rm th}\left[\Delta(\Gamma-\Gamma_0)\right]}
 {\Delta + \overline{y}_0 {\rm th}\left[\Delta(\Gamma-\Gamma_0)\right]}\nonumber\\
&=&|\Delta| \left( 1 + 2 \frac{\overline{y_0}-\Delta}{\overline{y}_0+\Delta}
\left({\Omega / \Omega_0} \right)^{2 \Delta}+\dots \right)\;,
\label{ysol}
\eeqn
where the solution goes through the point $y_0=\overline{y}_0$ at the reference
(log)energy cut-off, $\Gamma_0$. The second equation in Eq.(\ref{ysol}) is the approximate
form of the solution close to the line of fixed points, where in terms of the original
energy-scale variable $\Omega/\Omega_0 \ll 1$. We note that $y_0/\Delta$ is a unique
function of two dimensionless variables
\be
\frac{y_0}{\Delta}=y\left[\frac{\Delta}{\Delta_D}, \left(\frac{\Omega}{\Omega_D} \right)^{2 \delta}
 \right]\;,
\label{y_scal}
\ee
where $\Delta_D=\overline{y_0}$ and $\Omega_D=\Omega_0$.

In the following we relate the asymmetry parameter, $\Delta$, with the properties of the initial
distributions, $P_{\rm in}(h)$ and $R_{\rm in}(J)$. For this purpose we calculate first
the derivative:
\begin{eqnarray}
\frac{{\rm d}}{{\rm d} \Omega}\left[J^{\mu}\right]_{\rm av}=
R_0(\Omega,\Omega) \Omega^{\mu}+\int_0^{\Omega}\frac{{\rm d} R_0(J,\Omega)}{{\rm d} \Omega}
J^{\mu}{\rm d}J
\nonumber\\
=R_0(\Omega,\Omega) \Omega^{\mu} + (P_0(\Omega,\Omega)-R_0(\Omega,\Omega))\left[J^{\mu}\right]_{\rm av}
\nonumber\\
-
P_0(\Omega,\Omega) \int_0^{\Omega}{\rm d}JJ^{\mu} \int_J^{\Omega}{\rm d}J' R\left(\frac{J \Omega}
{J'} \Omega,\Omega\right) \frac{\Omega}{J'}
\;,
\label{dJmu1}
\end{eqnarray}
where in the second equation we have used the RG equation in Eq.(\ref{Rdiff}).
In the last term we change the order of the integration
and thus obtain:
\begin{equation}
\int_0^{\Omega}{\rm d}J' \int_0^{J'}{\rm d}J  J^{\mu}  R\left(\frac{J \Omega}
{J'} \Omega,\Omega\right) \frac{\Omega}{J'}=\Omega^{-\mu} \left[J^{\mu}\right]_{\rm av}^2
\;.
\label{dJmu2}
\end{equation}
One can evaluate in a similar way $({\rm d}/{\rm d} \Omega) [h^{-\mu}]_{\rm av}$ and then obtain
for the average value of the following derivative:
\begin{eqnarray}
\frac{{\rm d}}{{\rm d} \Omega}\left[\left(\frac{J}{h}\right)^{\mu}\right]_{\rm av}=
\left\{1-\left[\left(\frac{J}{h}\right)^{\mu}\right]_{\rm av}\right\} \times
\nonumber\\
\left\{P_0(\Omega,\Omega)\Omega^{-\mu} \left[J^{\mu}\right]_{\rm av}+
R_0(\Omega,\Omega)\Omega^{\mu} \left[h^{-\mu}\right]_{\rm av}\right\}
\;.
\label{dJhmu}
\end{eqnarray}
Notice that this quantity is vanishing for a parameter $\mu=\tilde{\mu}$, provided
$\left[(J/h)^{\tilde{\mu}}\right]_{\rm av}=1$,
thus $\tilde{\mu}$ defined in this way stays invariant along the RG trajectory. This relation
is valid  even if the starting RG steps are approximative. At the fixed point, $\Omega \to 0$,
with the solution in Eqs.(\ref{Psol})
and (\ref{Rsol}) we have evaluated the average:
\begin{equation}
\left[\left(\frac{J^2}{h^2}\right)^{\Delta}\right]_{\rm av}=1
\;.
\label{zeq}
\end{equation}
Consequently the asymmetry parameter, $\Delta$, is an invariant quantity of the RG transformation,
which is then determined by the initial distributions, as announced in Eq.(\ref{Delta}).
In the following we call $\Delta$ as the non-linear quantum control parameter of the RTIC.

Next we show that the RG equations in Eqs.(\ref{hdecimation}) and (\ref{Jdecimation}) become
asymptotically exact as the line of fixed points is approached, i.e., as $\Omega/\Omega_0 \to 0$.
Let us here consider the disordered Griffiths phase, $\Delta>0$, the
reasoning for $\Delta>0$ follows from duality. Here the ratio of
decimated bonds, $\Delta n_J$, and decimated transverse fields, $\Delta n_h$,
goes to zero as $\Delta n_J/\Delta n_h=R_0(\Omega,\Omega)/P_0(\Omega,\Omega)=
r_0/p_0 \sim \Omega^{2 \Delta}$, thus close to the fixed point almost exclusively
transverse fields are decimated out.
Then the probability, $Pr(\alpha)$, that the value of a coupling, $J$, being neighbor
to a decimated transverse field is $\Omega>J>\alpha \Omega$ with $0<\alpha<1$ is
given by
\begin{equation}
Pr(\alpha) \simeq \int_{\alpha\Omega}^{\Omega} R_0(J,\Omega) {\rm d} J =1-\alpha^{r_0}\approx
r_0 \ln(1/\alpha)\;,
\label{Pralpha}
\end{equation}
which goes to zero during iteration, since according to Eqs.(\ref{ysol}) and (\ref{y_0})
$r_0=R_0(\Omega,\Omega)\Omega \to 0$. Consequently the RG transformation becomes
asymptotically exact and the singularities, which are characterized by the
parameter $\Delta$ as calculated by the original distributions in Eq.(\ref{zeq}), are also exact.

\subsection{Relation between energy- and length-scale}

Next we are going to study the actual relation between the asymmetry or non-linear
quantum control parameter, $\Delta$, and the
Griffiths-McCoy singularities of the RTIC. For this we investigate the relation between the
energy scale, $\Omega$, and the
length scale, $L_{\Omega}$, by studying the fraction of non-decimated spins,
$n_{\Omega}$. When the energy
scale is decreased by an amount of ${\rm d}\Omega$ a fraction of spins.
${\rm d} n_{\Omega}= n_{\Omega}[P_0(\Omega,\Omega)+R_0(\Omega,\Omega)]$, is decimated
out, so that we obtain the differential equation:
\begin{equation}
\frac{{\rm d} n_{\Omega}}{{\rm d} \Omega}= n_{\Omega}[P_0(\Omega,\Omega)+R_0(\Omega,\Omega)]
\;,
\label{dnomega}
\end{equation}
what can be rewritten as
\begin{equation}
-\frac{{\rm d} \ln n_{\Omega}}{{\rm d} \ln \Omega}= -(r_0(\Omega)+p_0(\Omega))=-2y_0(\Omega)
\;.
\label{dnoGamma}
\end{equation}
Using the solution to $y_0(\Omega)$ in Eq.(\ref{ysol}) one can integrate Eq.(\ref{dnoGamma})
with the result:
\begin{equation}
n_{\Omega}=\left\{{\rm ch}\left[\Delta \ln\frac{\Omega_0}{\Omega}\right]+
\frac{\overline{y}_0}{\Delta} {\rm sh}\left[\Delta \ln\frac{\Omega_0}{\Omega}\right]
\right\}^{-2}
\;.
\label{nomega}
\end{equation}
To obtain the scaling form
at the critical point we take in Eq.(\ref{nomega}) the limits, $\Delta \to 0$ and
$\Gamma=-\ln \Omega \to \infty$, with however $\Delta \times \Gamma \to 0$ and obtain:
\begin{equation}
n_{\Omega}= \left[ 1+\overline{y}_0 \ln \frac{\Omega_0}{\Omega} \right]^{-2}
\sim \left[ \ln \frac{\Omega_0} {\Omega}\right]^{-2},\quad \Delta=0
\;,
\label{nomega0}
\end{equation}
which could be also directly calculated from Eq.(\ref{nomega}) with the critical
point solution in Eq.(\ref{sol0}). Thus, from Eq.(\ref{nomega0}) we get
for the typical distance between remaining spins, $L_{\Omega}$, as:
\begin{equation}
L_{\Omega}\sim \frac{1}{n_{\Omega}} \sim \left[ \ln \frac{\Omega_0}{\Omega}\right]^2,
\quad \Delta=0
\;,
\label{lomega0}
\end{equation}
which is just the relation in Eq.(\ref{scales}) as found earlier by Fisher\cite{fisher}.

In the Griffiths phases, $|\Delta|>0$, one obtains in Eq.(\ref{nomega}),
$n_{\Omega} \sim \Omega^{2|\Delta|}$,
in the limit $\Omega \to 0$. Consequently the relation between typical distance
between remaining spins, $L_{\Omega} \sim 1/n_{\Omega}$, and the energy scale is given by:
\begin{equation}
L_{\Omega}\simeq L_{\Omega_0} (\Delta+y_0)^2 \left( \frac{2}{\Delta} \right)^2
\left( \frac{\Omega_0}{\Omega}\right)^{2|\Delta|}
\sim \Omega^{-2 |\Delta|}
\;.
\label{lomega1}
\end{equation}
Thus $\Delta$ is simply related to the dynamical exponent, $z$,
\be
z=\frac{1}{2 |\Delta|}\;,
\ee
as announced in Eq.(\ref{z_exact}).

To obtain a relation between the non-linear quantum control parameter, $\Delta$. in Eq.(\ref{zeq}),
and the linear control parameter, $\delta$, as defined in Eq.(\ref{delta}) we perform a
Taylor-expansion for:
\begin{equation}
\left[J^{2\Delta}\right]_{\rm av}=1+2\Delta \left[\ln J\right]_{\rm av}+
2\Delta^2\left[\ln J)^2\right]_{\rm av}+O(\Delta^{3})
\;,
\label{Jexp}
\end{equation}
and similarly for $\left[h^{2\Delta}\right]_{\rm av}$. Putting these into Eq.(\ref{zeq})
we obtain, that $\Delta(\delta)=\delta+O(\delta^2)$, as announced in Eq.(\ref{Delta_delta}).

Closing this section we argue that the special solution in Eqs.(\ref{Psol}) and (\ref{Rsol})
is the true fixed-point solution as $\Omega \to 0$. First, we refer to Fisher's results
at the critical point\cite{fisher}, which justifies that  any non-singular
initial distribution is attracted by the special solution in Eqs.(\ref{Psol}) and (\ref{Rsol}).
Second, we consider a finite energy-scale, $\Omega>0$, when the non-asymptotic solution to
Eqs.(\ref{Pdiff}) and (\ref{Rdiff}) are given by the special solutions in Eqs.(\ref{Psol})
and (\ref{Rsol}) extended by non-universal functions, $P'_0(h,\Omega)$ and $R'_0(J,\Omega)$,
respectively. Inserting these non-asymptotic solutions into Eq.(\ref{Psol}) the relation in
Eq.(\ref{Rint1}) will be extended by other terms, containing $P'_0$ and $R'_0$. As $\Omega$
goes to zero, however, the second factor in Eq.(\ref{Rint1}) is diverging, therefore
the corrections become irrelevant and the relation in Eq.(\ref{Pedge}) will govern
the fixed-point behavior. Our third argument is based on numerical solutions of
Eqs.(\ref{Pdiff}) and (\ref{Rdiff}), which are evolving towards the special solutions in
Eqs.(\ref{Psol}) and (\ref{Rsol}) for different initial distributions.

Thus we can summarize that in the Griffiths phases the RTIC is uniquely characterized by
a non-universal quantum control parameter, $\Delta$, which is related to the dynamical exponent, $z$,
through Eq.(\ref{z_exact}).
The possible difference between two initial distributions leading to the same $\Delta$ is
given by the non-universal parameters, $\Omega_0$ and $\overline{y}_0$, which account
for the number of necessary RG steps until the fixed point distributions in
Eqs.(\ref{Psol}) and (\ref{Rsol}) are sufficiently approached.

\section{Renormalization of lengths and magnetic moments}

The scaling behavior of lengths and magnetic moments during renormalization
can be deduced from the joint distribution functions, $P_l(h,l,\Omega), R_l(J,l,\Omega)$
and $P_{\mu}(h,\mu,\Omega)$, as defined in Eqs.(\ref{distributions}). From here
on we drop the index $s$ or $b$ to indicate the type of the length.

\subsection{Scaling of lengths}

In the following we consider the joint distribution, $P_l(h,l,\Omega)$, and write down the relevant
evaluation equation when energy scale is lowered as $\Omega \to \Omega -{\rm d} \Omega$.
Generalizing the reasoning leading to Eq.(\ref{Pdiff}) we obtain:
\begin{eqnarray}
\frac{{\rm d} P_l(h,l,\Omega)}{{\rm d} \Omega}=P_l(h,l,\Omega)\left[R_0(\Omega,\Omega)-P_0(\Omega,\Omega)\right]
\nonumber\\
-\int_{h}^\Omega {\rm d} h_1 \frac{\Omega}{h_1} \int_0^{l} {\rm d} l_2 \int_0^{l-l_2} {\rm d} l_1
R_l(\Omega,l_2,\Omega)  P_l(h_1,l_1,\Omega)
\nonumber\\
\times P_l\left(\frac{h}{h_1}\Omega,l-l_1-l_2,\Omega\right) 
\;,
\label{h-l-diff}
\end{eqnarray}
and similarly for the coupling distribution by interchanging $R \leftrightarrow P$ and
$J_i \leftrightarrow h_i$, as in Eqs.(\ref{Pdiff}) and (\ref{Rdiff}). The second term in
the r.h.s. of Eq.(\ref{h-l-diff}) can be written as a convolution in terms of the variable
$l'=l_1+l_3=l-l_2$ as
\begin{eqnarray}
-\int_0^{l} {\rm d} l' R_l(\Omega,l-l',\Omega) \int_0^{l'} {\rm d} l_1
\nonumber\\
\times
\int_{h}^\Omega {\rm d} h_1 \frac{\Omega}{h_1}  
  P_l(h_1,l_1,\Omega) P_l(\frac{h}{h_1}\Omega,l'-l_1,\Omega) 
\;.
\label{h-l-diff1}
\end{eqnarray}
Consequently taking the Laplace-transform:
\begin{eqnarray}
\tilde{P}_l(h,\lambda,\Omega)&=&\int_0^{\infty} {\rm e}^{-l \lambda} P_l(h,l,\Omega) {\rm d} l
\nonumber\\
\tilde{R}_l(J,\lambda,\Omega)&=&\int_0^{\infty} {\rm e}^{-l \lambda} R_l(J,l,\Omega) {\rm d} l\;,
\label{lapl-tr}
\end{eqnarray}
we obtain a simpler relation
\begin{eqnarray}
&.&\frac{{\rm d} \tilde{P}_l(h,\lambda,\Omega)}{{\rm d} \Omega}=\tilde{P}_l(h,\lambda,\Omega)
\left[\tilde{R}_l(\Omega,0,\Omega)-\tilde{P}_l(\Omega,\lambda,\Omega)\right]
\nonumber\\
&-&\tilde{R}_l(\Omega,\lambda,\Omega) \int_{h}^\Omega {\rm d} h' \tilde{P}_l(h',\lambda,\Omega)
\tilde{P}_l(\frac{h}{h'}\Omega,\lambda,\Omega) \frac{\Omega}{h'}
\;,
\label{P-lapl-diff}
\end{eqnarray}
and similarly for the coupling distribution:
\begin{eqnarray}
&.&\frac{{\rm d} \tilde{R}_l(J,\lambda,\Omega)}{{\rm d} \Omega}=\tilde{R}_l(J,\lambda,\Omega)
\left[\tilde{P}_l(\Omega,0,\Omega)-\tilde{R}_l(\Omega,\lambda,\Omega)\right]
\nonumber\\
&-&\tilde{P}_l(\Omega,\lambda,\Omega) \int_{J}^\Omega {\rm d} J' \tilde{R}_l(J',\lambda,\Omega)
\tilde{R}_l(\frac{J}{J'}\Omega,\lambda,\Omega) \frac{\Omega}{J'}
\;.
\label{R-lapl-diff}
\end{eqnarray}
Note that the different $\lambda$ components are separated, which makes possible to
solve the equations. For $\lambda=0$, when $\tilde{P}_l(h,0,\Omega)=P_0(h,\Omega)$
and $\tilde{R}_l(J,0,\Omega)=R_0(J,\Omega)$, the solutions
are given in Eqs.(\ref{Psol}) and (\ref{Rsol}). With this guidance
we are looking for the solution for general $\lambda$ in the form:
\begin{eqnarray}
\tilde{P}_l(h,\lambda,\Omega)&=&\frac{\pi_l(\lambda,\Omega)}{\Omega}
\left(\frac{\Omega}{h}\right)^{1-p_l(\lambda,\Omega)}
\label{P-lamb-sol}\\
\tilde{R}_l(J,\lambda,\Omega)&=&\frac{\rho_l(\lambda,\Omega)}{\Omega}
\left(\frac{\Omega}{J}\right)^{1-r_l(\lambda,\Omega)}\;,
\label{R-lamb-sol}
\end{eqnarray}
where now $p_l(0,\Omega)=\pi_l(0,\Omega)=p_0(\Omega)$ and $r_l(0,\Omega)=\rho_l(0,\Omega)=r_0(\Omega)$,
whereas for $\lambda>0$
$p_l(\lambda,\Omega)>\pi_l(\lambda,\Omega)$ and $r_l(\lambda,\Omega)>\rho_l(\lambda,\Omega)$.
This latter relation follows from the fact that the average length of a bond,
$\overline{l}_b>0$, and that of a spin cluster, $\overline{l}_s>0$, is given by:
\begin{equation}
\overline{l}_b=\lim_{\lambda \to 0} \frac{1}{\lambda}\left[1-\frac{\rho_l(\lambda,\Omega)}
{r_l(\lambda,\Omega)}\right],\quad \overline{l}_s=\lim_{\lambda \to 0} \frac{1}
{\lambda}\left[1-\frac{\pi_l(\lambda,\Omega)} {p_l(\lambda,\Omega)}\right]\;.
\label{lengths}
\end{equation}
Inserting the functions in Eqs.(\ref{P-lamb-sol}) and (\ref{R-lamb-sol}) into Eqs.(\ref{P-lapl-diff})
and (\ref{R-lapl-diff} we obtain a set of ordinary differential equations:
\begin{eqnarray}
\frac{{\rm d} p_l}{{\rm d} \Gamma}&=&-\pi_l \rho_l,\quad
\frac{{\rm d} \pi_l}{{\rm d} \Gamma}=-\pi_l(p_l-\pi_l+\rho_l)
\nonumber\\
\frac{{\rm d} r_l}{{\rm d} \Gamma}&=&-\pi_l \rho_l,\quad
\frac{{\rm d} \rho_l}{{\rm d} \Gamma}=-\rho_l(r_l-\rho_l+\pi_l)
\;.
\label{pi-p-rho-r}
\end{eqnarray}
involving the functions $p_l,\pi_l,r_l$ and $\rho_l$.

These equations are solved in the Appendix. Here we only consider the scaling behavior of
the average lengths and for this it is enough to treat the small $\lambda$ expansions
up to linear order:
\begin{eqnarray}
p_l(\lambda,\Omega)&=&p_0(\Omega)+\lambda p_1(\Omega),\quad
\pi_l(\lambda,\Omega)=p_0(\Omega)+\lambda \pi_1(\Omega)
\nonumber\\
r_l(\lambda,\Omega)&=&r_0(\Omega)+\lambda r_1(\Omega),\quad
\rho_l(\lambda,\Omega)=r_0(\Omega)+\lambda \rho_1(\Omega)
\;,
\nonumber\\
\label{pi-p-rho-r_corr}
\end{eqnarray}
Inserting the expressions in Eqs.(\ref{pi-p-rho-r_corr}) into Eq.(\ref{pi-p-rho-r}) we
obtain for the correction terms:
\begin{eqnarray}
\frac{{\rm d} p_1}{{\rm d} \Gamma}&=&-\pi_1 r_0-\rho_1 p_0 ,\quad
\frac{{\rm d} \pi_1}{{\rm d} \Gamma}=-\pi_1 r_0-p_0(p_1-\pi_1+\rho_1)
\nonumber\\
\frac{{\rm d} r_1}{{\rm d} \Gamma}&=&-\pi_1 r_0-\rho_1 p_0,\quad
\frac{{\rm d} \rho_1}{{\rm d} \Gamma}=-\rho_1 p_0-\rho_0(r_1-\rho_1+\pi_1)
\;.
\nonumber\\
\label{pi-p-rho-r_corr1}
\end{eqnarray}
Now noticing that
\be
\frac{{\rm d} \ln (p_1-\pi_1)}{{\rm d} \Gamma}=p_0,\quad
\frac{{\rm d} \ln (r_1-\rho_1)}{{\rm d} \Gamma}=r_0\;,
\label{p-pi}
\ee
we obtain after integration using Eqs.(\ref{Pedge}) and (\ref{Redge}), that
$p_1-\pi_1=A_p/r_0$ and $r_1-\rho_1=A_r/p_0$, where $A_p,A_r$ are integration constants.
Consequently the average lengths from Eq.(\ref{lengths}) are given by:
\be
\overline{l}_s=\frac{p_1-\pi_1}{p_0}=\overline{l}_s(\Omega_0)\frac{r_0(\Omega_0)p_0(\Omega_0)}
{r_0(\Omega)p_0(\Omega)}=\overline{l}_s(\Omega_0)\frac{\overline{y}_0^2-\Delta^2}
{{y}_0^2-\Delta^2}\;,
\label{l_s_av}
\ee
and
\be
\overline{l}_b=\frac{r_1-\rho_1}{r_0}=\overline{l}_b(\Omega_0)\frac{r_0(\Omega_0)p_0(\Omega_0)}
{r_0(\Omega)p_0(\Omega)}=\overline{l}_b(\Omega_0)\frac{\overline{y}_0^2-\Delta^2}
{{y}_0^2-\Delta^2}\;.
\label{l_b_av}
\ee
At the line of fixed points, $\Omega \to 0$, one can see that
$\overline{l}_s \sim \overline{l}_b \sim L_{\Omega}$,
consequently the previous interpretation of the dynamical exponent, $z$, in Eq.(\ref{lomega1}) is
justified also with the average lengths-scales.

Now to calculate the correlation length. $\xi$, in the paramagnetic phase, $\Delta>0$, one should take
into account that the ratio of (non-decimated transverse-fields)/(non-decimated couplings)
at an energy-scale,
$\Omega$, is given by $p_0/r_0$. Consequently the number of non-decimated spins in a cluster is given by
$\sim \overline{l}_s p_0/r_0 \sim 1/r_0^2 \sim \Delta^{-2}$, which stays constant as  the energy-scale
is lowered. This quantity is actually the measure of the size of the average correlated domain in
the system, where the couplings
between the spins, being larger then the transverse fields, are decimated out. Therefore in this way we
have an estimate for the correlation length close to the critical point:
\be
\xi \sim \Delta^{-2} \sim \delta^{-2}\;,
\ee
which is consistent with Fisher's result in Eq.(\ref{nu}).

\subsection{Scaling of magnetization moments}

In this subsection we perform a similar calculation about the joint distribution function,
$P_{\mu}(h,\mu,\Omega)$, and calculate the average size of a magnetic moment,
$\overline{\mu}(\Omega)$, as a function of the energy cut-off. The joint distribution function,
$P_{\mu}(h,\mu,\Omega)$, satisfies the differential equation:
\begin{eqnarray}
&.&\frac{{\rm d} P_{\mu}(h,\mu,\Omega)}{{\rm d} \Omega}=P_{\mu}(h,\mu,\Omega)
\left[R_0(\Omega,\Omega)-P_0(\Omega,\Omega)\right]-
\nonumber\\
&R&_0(\Omega,\Omega)\int_{h}^\Omega {\rm d} h' \frac{\Omega}{h'} \int_0^{\mu} {\rm d} \mu' 
P_{\mu}(h',\mu',\Omega)P(\frac{h}{h'}\Omega,\mu-\mu',\Omega) 
\;,
\nonumber\\
\label{h-mu-diff}
\end{eqnarray}
what can be derived along the lines of Eqs.(\ref{Pdiff}) and (\ref{h-l-diff}).
The second term in the r.h.s. of Eq.(\ref{h-mu-diff}) is a convolution therefore
we introduce the Laplace-transform:
\begin{equation}
\tilde{P}_{\mu}(h,s,\Omega)=\int_0^{\infty} {\rm e}^{-\mu s} P_{\mu}(h,\mu,\Omega) {\rm d} \mu
\;,
\label{lapl-tr-s}
\end{equation}
which satisfies the relation
\begin{eqnarray}
&.&\frac{{\rm d} \tilde{P}_{\mu}(h,s,\Omega)}{{\rm d} \Omega}=\tilde{P}_{\mu}(h,s,\Omega)
\left[R_0(\Omega,\Omega)-\tilde{P}_{\mu}(\Omega,0,\Omega)\right]
\nonumber\\
&-&R_0(\Omega,\Omega) \int_{h}^\Omega {\rm d} h' \tilde{P}_{\mu}(h',s,\Omega)
\tilde{P}_{\mu}(\frac{h}{h'}\Omega,s,\Omega) \frac{\Omega}{h'}
\;.
\label{P-s-diff}
\end{eqnarray}
In Eq.(\ref{P-s-diff}) the different $s$ components are separated, for $s=0$, when
$\tilde{P}_{\mu}(h,0,\Omega)=P_0(h,\Omega)$ the solution
is given in Eqs.(\ref{Psol}). As for the joint distribution of the lengths, $P_l(h,\lambda,\Omega)$,
in Eq.(\ref{P-lamb-sol})
we are looking for the solution for general $s$ in the form:
\begin{equation}
\tilde{P}_{\mu}(h,s,\Omega)=\frac{\pi_{\mu}(s,\Omega)}{\Omega}
\left(\frac{\Omega}{h}\right)^{1-p_{\mu}(s,\Omega)}
\;.
\label{P-s-sol}
\end{equation}
Here again $p_{\mu}(0,\Omega)=\pi_{\mu}(0,\Omega)=p_0(\Omega)$,
whereas $p_{\mu}(s,\Omega)>\pi_{\mu}(s,\Omega)$ for $s>0$, since the average cluster moment,
$\overline{\mu}>0$, is given by:
\begin{equation}
\overline{\mu}=\lim_{s \to 0} \frac{1}{s}\left[1-\frac{\pi_{\mu}(s,\Omega)}
{p_{\mu}(s,\Omega)}\right]\;.
\label{moment}
\end{equation}
Putting Eq.(\ref{P-s-sol}) into Eq.(\ref{h-mu-diff}) we find that the functions,
$p_{\mu}$ and $\pi_{\mu}$, satisfy the differential equations:
\begin{eqnarray}
\frac{{\rm d} p_{\mu}}{{\rm d} \Gamma}&=&-\pi_{\mu} r_0\\
\nonumber
\frac{{\rm d} \pi_{\mu}}{{\rm d} \Gamma}&=&-\pi_{\mu} (r_0-p_0+p_{\mu})\;.
\label{p-pi-tilde}
\end{eqnarray}
Keeping in mind that the average cluster moment, $\overline{\mu}$, and thus the average
magnetization, $m$, defined as:
\begin{equation}
m=\frac{\overline{\mu}}{\overline{l}_s}\;,
\label{magn}
\end{equation}
is related to the small $s$ asymptotics of the distribution in Eq.(\ref{P-s-sol}),
we perform the expansions up to linear order:
\begin{equation}
p_{\mu}(s,\Omega)=p_0(\Omega)+s\tilde{p}_1(\Omega),\quad
\pi_{\mu}(s,\Omega)=p_0(\Omega)+s\tilde{\pi}_1(\Omega)\;.
\label{p-pi-exp}
\end{equation}
For the correction terms, $\tilde{p}_1$ and  $\tilde{\pi}_1$, we derive differential
equations in terms of the functions $p_0$ and $r_0$ as:
\begin{eqnarray}
\frac{{\rm d} \tilde{p}_1}{{\rm d} p_0}&=&\frac{\tilde{\pi}_1}{p_0}\\
\nonumber
\frac{{\rm d} \tilde{\pi}_1}{{\rm d} p_0}&=&\frac{\tilde{\pi}_1}{p_0}+\frac{\tilde{p}_1}{r_0}\;,
\label{p1-pi1}
\end{eqnarray}
which leads to:
\begin{equation}
(y_0^2-\Delta^2)\frac{{\rm d}^2 \tilde{p}_1}{{\rm d} y_0^2}=\tilde{p}_1\;,
\label{p1-y}
\end{equation}
where $y_0=y_0(\Omega)$ is given in Eq.(\ref{ysol}). We note that with the solution for $\tilde{p}_1$
we have for the average cluster moment:
\begin{equation}
\overline{\mu}=\frac{\tilde{p}_1-\tilde{\pi}_1}{p_0}=
-\frac{\int_{\overline{y}_0}^{y_0} {\rm d} y_0^{'} \tilde{p}_1(y_0^{'})/(y_0^{'}-\Delta)}{y_0+\Delta}\;.
\label{mu-y}
\end{equation}
{\it At the critical point} with $\Delta=0$ the solution of Eq.(\ref{p1-y}) is given in a
simple power-law form:
\begin{equation}
\tilde{p}_1=y_0^{-\tau},\quad \Delta=0\;,
\label{pysol0}
\end{equation}
where $\tau=(\sqrt{5}-1)/2$ is the positive root of the equation: $\tau(\tau+1)=1$. (The
other linearly independent solution with $\tau=-(\sqrt{5}+1)/2$ is physically not
acceptable, since the average cluster moment would be smaller than 1.) From Eq.(\ref{p1-pi1})
we have $\tilde{\pi}_1=-\tau y_0^{-\tau}$ and
using Eq.(\ref{mu-y}) we obtain for the average cluster moment at the critical
point:
\begin{equation}
\overline{\mu}={\rm const}~y_0^{-(1+\tau)} = \overline{\mu}_0 \left[ \ln
\left(\frac{\Omega_0}{\Omega}\right)\right]^{\Phi},~~
\Phi=\frac{1}{\tau}=\frac{1+\sqrt{5}}{2}\;.
\label{Phi}
\end{equation}
In this way we have re-derived Fisher's result\cite{fisher} about the scaling
behavior of the average cluster moment in a direct way.

{In the Griffiths phases} with $\Delta \ne 0$ the differential equation in Eq.(\ref{p1-y})
in terms of the variable $y=y_0/\Delta$
is related to the Legendre differential equation and the physically acceptable
solution can be expressed by the hypergeometric function\cite{AS}, $F(a,b;c;z)$, as
\begin{equation}
\tilde{p}_1= |\Delta|^{-\tau} y^{-\tau}F\left(\frac{\tau}{2},
\frac{1}{2}+\frac{\tau}{2};\frac{3}{2}+\tau;
\frac{1}{y^2}\right)=|\Delta|^{-\tau} f_1(y)\;,
\label{pysol}
\end{equation}
where, in the limit $\Delta \to 0$ we recover the solution at the critical point in
Eq.(\ref{pysol0}).
From Eq.(\ref{p1-pi1}) we obtain
\begin{eqnarray}
\tilde{\pi}_1&=&-|\Delta|^{-\tau}(y-1) y^{-(\tau+1) }F\left(\frac{\tau}{2}+1,
\frac{1}{2}+\frac{\tau}{2};\frac{3}{2}+\tau;
\frac{1}{y^2}\right)
\nonumber\\
&=&|\Delta|^{-\tau} \phi_1(y)\;,
\label{piysol}
\end{eqnarray}
and putting Eqs.(\ref{pysol}) and (\ref{piysol}) into Eq.(\ref{mu-y})
we obtain for the average cluster moment
\begin{equation}
\overline{\mu}={\rm const}|\Delta|^{-\tau-1} \frac{f(y)}{y+1}\;
\label{mu_o}
\end{equation}
where $f(y)=f_1(y)-\phi_1(y)$. Here one should differentiate between the paramagnetic
($\Delta>0,~y>0$) and the ferromagnetic ($\Delta>0,~y>0$) phases. In the former case the
average cluster moment is approaching a finite limiting value, as $\Omega/\Omega_0 \to 0$,
whereas in the ferromagnetic phase, where $y \to 1^{-}$ in the fixed point, thus
$\overline{\mu}$ is divergent, as $\overline{\mu}(\Omega) \sim \Omega^{-2 |\Delta|}$.
For the average magnetization in
Eq.(\ref{magn}) then one obtains:
\begin{equation}
m=m_0 \frac{(1-y) f(y)}{(1-\overline{y}) f(\overline{y})}\;,
\label{magn-f}
\end{equation}
where $m_0$ is the average magnetization at $\Omega=\Omega_0$  and $\overline{y}$ denotes the
value of the variable $y$ at the same energy-scale. The average magnetization in the
paramagnetic phase is zero, whereas in the ferromagnetic phase one have to
evaluate  Eq.(\ref{magn-f})
along the line of semi-critical fixed points, $\Omega/\Omega_0 \to 0$. Here taking the limit
$|\Delta| \ll 1$, i.e. being close to the critical point we have $(1-\overline{y})^{-1} \sim |\Delta|$
and $f(\overline{y}) \sim |\Delta|^{\tau}$ so that\cite{remark}:
\begin{equation}
m={\rm const} |\Delta|^{1-\tau}={\rm const} |\delta|^{1-\tau}\;.
\label{magn-d}
\end{equation}
From Eq.(\ref{magn-d}) one can read the critical exponent of the average magnetization as:
\begin{equation}
\beta=1-\tau=2-\Phi\;,
\label{betan}
\end{equation}
which corresponds to Fisher's result in Eq.(\ref{beta}).

\section{Scaling of thermodynamical quantities}

In the previous Sections we have presented the solution of the RG equations in the
entire Griffiths region for the
distribution of couplings, transverse fields, lengths
and magnetization moments. Then, with those distributions, average quantities, such as length-scales,
magnetization, etc. are calculated at zero temperature and in the absence of a longitudinal
magnetic field. In this Section we extend these calculations and determine the scaling form
of singular thermodynamic quantities as a function of a small, but finite temperature,
$T>0$, or magnetic field, $H>0$.

To treat the effect of a small finite temperature in the RG scheme one should first notice
that the thermal energy sets in an energy scale, $\Omega_T \sim T$, and the RG decimation
should be stopped as $\Omega$ is lowered to $\Omega_T$.
At that energy scale a fraction of spin clusters, $n_{\Omega_T}$, in Eq.(\ref{nomega})
is not decimated out and these spins are loosely coupled comparing with the temperature, $T$.
Consequently the entropy per spin, $s$, is given as the contribution of non-interacting spin
clusters:
\be
s \simeq  n_{\Omega_T} \ln 2 \;,
\label{entropy}
\ee
whereas the specific heat can be obtained through derivation: $c_V=T(\partial s/\partial T)$.
From Eqs.(\ref{entropy}) and (\ref{nomega}) we obtain for the singular behavior:
\be
s(T) \sim c_V(T) \sim T^{2 |\Delta|}\;,
\label{entropy_d}
\ee
which is valid both in the ordered and in the disordered Griffiths phases.

Next, we consider the effect of a small longitudinal field, $H>0$, at zero temperature.
During renormalization the local longitudinal field, $H_l$, at site $l$ is transformed as
\be
\tilde{H}_l=H \mu_l\;,
\ee
so that the energy-scale related to the longitudinal field is given by
$\Omega_H=H \overline{\mu}(\Omega)$. As $\Omega$ is lowered to $\Omega_H$, i.e. when
the energy scale satisfies the equation
\be
\Omega_H=H \overline{\mu}(\Omega_H)\;,
\ee
the RG procedure is stopped and the remaining spin clusters are practically uncoupled.
Then the average magnetization and the average susceptibility satisfy the equations:
\be
m(H)=m(\Omega=\Omega_H),\quad \chi=\frac{\partial m}{\partial H}\;.
\ee
In the {\it disordered Griffiths phase}, where $\overline{\mu}(\Omega_H)$ has a $\Omega_H$ independent
limiting value, we have $\Omega_H \sim H$, consequently from Eq.(\ref{magn-f}) the
singular behavior is given by
\be
m(H) \sim \left(\frac{H}{H_D}\right)^{2\Delta},\quad \Delta>0\;.
\ee
More generally the scaling form is given in Eq.(\ref{m_scaling}), where the
scaling function can be computed using Eqs.(\ref{magn-f}), (\ref{ysol}) and (\ref{y_scal}).
Similarly one obtains for the scaling of the susceptibility in the disordered Griffiths phase:
\be
\chi(H) \sim \left(\frac{H}{H_D}\right)^{-1+2\Delta},\quad 
\chi(T) \sim T^{-1+2\Delta},\quad
\Delta>0\;,
\ee
where the temperature dependence follows from the scaling relation, $\Omega_H \sim \Omega_T$.

In the {\it ordered Griffiths phase}, where $\overline{\mu}(\Omega_H) \sim \Omega_H^{-2 |\Delta|}$,
as given above Eq.(\ref{magn-f}) we have
$\Omega_H \sim H^{1/(1+2 |\Delta|)}$. Putting this result into Eq.(\ref{magn-f}) and using the
asymptotic expansion for the hypergeometric functions\cite{AS} in Eqs.(\ref{pysol}) and
(\ref{piysol}) we obtain for
the leading field dependence of the magnetization:
\be
m(H)-m(0) \sim \left(\frac{H}{H_D}\right)^{\frac{2|\Delta|}{1+2|\Delta|}} \ln\left(\frac{H}{H_D}\right),
\quad \Delta<0\;,
\ee
and similarly for the susceptibility:
\be
\chi(H) \sim \left(\frac{H}{H_D}\right)^{\frac{-1}{1+2|\Delta|}} \ln\left(\frac{H}{H_D}\right),
\quad \Delta<0\;.
\ee
Note that in the ordered Griffiths phase the singularity exponent is different from that in
the disordered Griffiths phase and there is a logarithmic correction term. The temperature
dependence of the susceptibility, which follows from the relation $\Omega_H \sim \Omega_T$,
is given by:
\be
\chi(T) \sim T^{-1+2|\Delta|} \ln T,\quad
\Delta<0\;.
\ee
We can conclude this Section that all the singularities of different physical
quantities, both in the (strongly) ordered and disordered Griffiths phases can
be expressed by the non-linear quantum control parameter, $\Delta$.

\section{Discussion}

In this paper the strong disorder RG method is applied in the strongly disordered and
strongly ordered Griffiths phases of the random transverse-field Ising spin chain. With
this calculation we have demonstrated that the RG method leads to asymptotically
exact results in the entire Griffiths region. The key concept of our solution is the
introduction of a non-linear quantum control parameter, $\Delta$, which stays invariant
under the RG transformation, even if the renormalization is approximative in the
starting decimation steps. $\Delta$, which is a measure of the asymmetry in the renormalization
between the couplings and the transverse fields, is simply related to the dynamical exponent,
$z$, and all the singularities of the different physical quantities in the Griffiths phases can
be expressed with it. In this way we have presented an example for an RG transformation, where
the relevant {\it non-linear scaling field}\cite{wegner} outside the critical fixed point is exactly
constructed and the off-critical singularities are analytically calculated.
The line of fixed points controlling the singular behavior in the
Griffiths phases are found {\it strongly attractive}: for any weak initial disorder, having
the same asymmetry parameter, $\Delta$, the system scales into the same fixed point. This
is a remarkable weak-universality property of the system. We note that previous
numerical\cite{bigpaper,numerical} and analytical\cite{analytical} results about the RTIC are
in accordance with our RG findings.

At this point one may ask the question, how far these results are general and could
apply for other random quantum spin systems. The above scenario is certainly valid for those
problems, which can be mapped to the RTIC, so that the RG equations can be transformed
into an equivalent form. Free fermionic spin-$1/2$ models, such as the random $XX$ model
with dimerization or the random $XY$ model with an $X/Y$ anisotropy are such
examples\cite{fisherxx,xx}.
Also the one-dimensional Sinai-walk problem, i.e. random walk in a random environment
with a global bias\cite{sinai}, can be mapped to the RTIC\cite{itr99},
thus a renormalization analysis\cite{fldm} leads
to asymptotically exact results for this problem, too.

Other, more general random quantum spin systems could belong into two main classes:
i) systems having an IRFP and a line of semi-critical fixed points which are
{\it strongly attractive}, and ii) those models where a cross-over phenomena takes
place when the strength of disorder is increased. Models belonging to the
first class are, among others, the random spin-$1/2$ Heisenberg chain\cite{fisherxx} and the
random $q \ge 2$ states quantum Potts chains\cite{qPotts}. For these systems we expect that
a strong disorder RG-calculation leads to asymptotically exact results, too,
and the physical picture obtained in the analytical treatment of
the RTIC stays qualitatively correct. Indeed scaling arguments and numerical
calculations about specific models are in favor of our conjecture\cite{ijl01}.

Into the second class of models belong, among others, the random
quantum clock and quantum Ashkin-Teller chains\cite{cli01} and probably several higher dimensional
systems (two-dimensional random transverse-field Ising model\cite{2drg,lkir}, etc.). For these
problems
the disorder should exceed a limiting strength, when the IRFP and the line of semi-critical
fixed points become attractive. Above this limiting disorder value the strong disorder
RG could be asymptotically exact. To verify this
possibility, however, one should perform detailed numerical investigations.

Acknowledgment: The author is grateful to R. Juh\'asz and P. Lajk\'o  for
previous cooperation in the subject and to J-C. Angl\'es d'Auriac, E. Carlon,
R. Melin, H. Rieger and L. Turban for useful discussions. This work has been
supported by a German-Hungarian exchange program (DAAD-M\"OB), by the Hungarian
National Research Fund under grant No OTKA
TO23642, TO25139, TO34183  and  MO28418, by the Ministry of Education under grant
No. FKFP 87/2001 and by the Centre of Excellence ICA1-CT-2000-70029.

\appendix
\section*{Distribution function of lengths}

Here we present the general solution for the joint distribution functions, $P_l(h,l,\Omega)$ and
$R_l(J,l,\Omega)$, in the vicinity of the line of fixed points, $\Omega/\Omega_0 \to 0$.
This amounts to
solve the set of differential equations in Eqs.(\ref{pi-p-rho-r}) involving the functions
$p_l(\lambda,\Omega),\pi_l(\lambda,\Omega),r_l(\lambda,\Omega)$ and $\rho_l(\lambda,\Omega)$,
which appear in the Laplace-transforms, $\tilde{P}_l(h,\lambda,\Omega)$ and
$\tilde{R}_l(h,\lambda,\Omega)$, in Eqs.(\ref{P-lamb-sol}) and (\ref{R-lamb-sol}).

We start with the solution {\it at the critical point} where the fixed point distribution
of the couplings and the transverse fields are identical, so that
$p_l(\lambda,\Omega)=r_l(\lambda,\Omega)$
and $\pi_l(\lambda,\Omega)=\rho_l(\lambda,\Omega)$. Here we have just two differential
equations:
\begin{equation}
\frac{{\rm d} p_l}{{\rm d}\Gamma}=-\pi_l^2,\quad
\frac{{\rm d} \ln \pi_l}{{\rm d} \Gamma}=-p_l \pi_l\;.
\label{pi-p}
\end{equation}
From Eq.(\ref{pi-p}) follows that  ${\rm d}(p_l^2-\pi_l^2)/{\rm d}\Gamma=0$, therefore
$p_l^2$ and $\pi_l^2$ differ by an $\Omega$-independent term:
\begin{equation}
\pi_l^2(\lambda,\Omega)=p_l^2(\lambda,\Omega)-c^2(\lambda)\;,
\label{pi-p-C}
\end{equation}
and we can write a simple differential equation:
\begin{equation}
\frac{{\rm d} p_l}{{\rm d} \Gamma}+p_l^2=c^2\;.
\label{p-C}
\end{equation}
Since Eq.(\ref{p-C}) is equivalent to Eq.(\ref{dif3}), we have for its solution from Eq.(\ref{ysol})
with the substitution $\Delta \to c$
\begin{equation}
p_l=\frac{p_l^0 c+{c^2} {\rm th}\left[c \ln(\Omega_0/\Omega)\right]}
{c+p_l^0 {\rm th}\left[c \ln(\Omega_0/\Omega)\right]}\;,
\label{psol}
\end{equation}
where now $p_l^0=p(\lambda,\Omega_0)$ at a reference point, $\Omega=\Omega_0$.

Close to the line of fixed points, as $\Omega_0/\Omega \to 0$ we should have $c(\lambda) \to 0$,
in order to have a finite scaling combination in Eq.(\ref{psol}). In this
small $\lambda$ limit we obtain from Eqs.(\ref{lengths}) and (\ref{pi-p-C}) that
$c^2(\lambda) = a^2 \lambda + O(\lambda^2)$ where the value of the constant
$a$ is connected to the average lengths at $\Omega=\Omega_0$, $\overline{l}_0=
\overline{l}(\Omega_0)$. as $\overline{l}_0=a^2/2(p_l^0)^2$. As $\Omega \to 0$ the average
length of a cluster or bond is divergent as:
\begin{equation}
\overline{l}=\overline{l}_0  p_0^2 \left( \ln \frac{\Omega_0}{\Omega}\right)^2\;,
\label{l-scale}
\end{equation}
which is the same as the typical distance between remaining spins, $L_{\Omega}$, as
given in Eq.(\ref{lomega0}) if we make the identification, $a^2=p_0$. To obtain the
joint distribution of the fields (couplings) and lengths we use the fact that at the fixed point
of the RG transformation the appropriate scaling variable in Eq.(\ref{psol}) is
$a\lambda^{1/2} \ln(\Omega_0/\Omega)=O(1)$, therefore:
\begin{eqnarray}
p_l(\lambda,\Omega)&=&a\lambda^{1/2} {\rm cth}\left[ a\lambda^{1/2}
\left(\ln \frac{\Omega_0}{\Omega}\right) \right]
\nonumber\\
\pi_l(\lambda,\Omega)&=&a\lambda^{1/2} {\rm sh}^{-1}\left[ a\lambda^{1/2}
\left(\ln \frac{\Omega_0}{\Omega}\right) \right]\;.
\label{p-pi-sol}
\end{eqnarray}
Consequently at the fixed point $P_l(h,l,\Omega)$ and $R_l(J,l,\Omega)$ can be obtained
by the inverse Laplace-transform of Eqs.(\ref{P-lamb-sol}) and (\ref{R-lamb-sol}) with
Eqs.(\ref{p-pi-sol}).

In the Griffiths region, i.e. {\it outside the critical point} one should
consider the system of four coupled differential equations in Eqs.(\ref{pi-p-rho-r}),
where one can construct two $\Omega$-independent combinations of the variables:
\begin{equation}
p_l(\lambda,\Omega)-r_l(\lambda,\Omega)=2\Delta(\lambda)\;,
\label{p-r}
\end{equation}
and
\begin{equation}
p_l(\lambda,\Omega)r_l(\lambda,\Omega)-\pi_l(\lambda,\Omega)\rho_l(\lambda,\Omega)=D(\lambda)^2\;.
\label{pr-pirho}
\end{equation}
Thus there are two variables left: $y_l=(p_l+r_l)/2$ and $u_l=p_l-\pi_l$, which satisfy the
differential equations:
\begin{equation}
\frac{{\rm d} y_l}{{\rm d} \Gamma} + y_l^2=d^2,\quad d^2=\Delta(\lambda)^2 + D(\lambda)^2\;,
\label{DEy}
\end{equation}
and
\begin{equation}
\frac{{\rm d} u_l}{{\rm d} \Gamma} + u_l^2-\left(y_l+|\Delta(\lambda)|\right)u_l=0\;.
\label{DEu}
\end{equation}
The solution of Eq.(\ref{DEy}) is analogous to that of Eq.(\ref{p-C}) and immediately
given by:
\begin{equation}
y_l=\frac{y_l^0 d+{d^2} {\rm th}\left[d (\Gamma-\Gamma_0)\right]}
{d+y_l^0 {\rm th}\left[d (\Gamma-\Gamma_0) \right]}\;,
\label{DEy_sol}
\end{equation}
where $y_l^0=y_l(\lambda,\Omega_0)$ at the reference energy, $\Omega=\Omega_0$.

To integrate Eq.(\ref{DEu}) we first notice that it is a Bernoulli-type differential
equation and its solution can be expressed as:
\begin{equation}
\frac{1}{u_l}=E(\Gamma) \left(\int\frac{1}{E(\Gamma')} {\rm d} \Gamma' + C\right)\;,
\label{1/u}
\end{equation}
with
\begin{equation}
E(\Gamma)=\exp \left(-\int [y_l(\Gamma')+|\Delta(\lambda)|] {\rm d} \Gamma' \right)\;.
\end{equation}
Using the solution for $y_l(\Gamma)$ in Eq.(\ref{DEy_sol}) we can perform the integration
for $E(\Gamma)$
as
\begin{equation}
E(\Gamma)=e^{-\Gamma |\Delta| } \left\{ d~{\rm ch}[d~(\Gamma-\Gamma_0)]+
y_l^0~{\rm sh}[d~(\Gamma-\Gamma_0)] \right\}^{-1}\;,
\end{equation}
and putting it into Eq.(\ref{1/u}) one can integrate once more giving:
\begin{equation}
u_l=\frac{f_+(\Gamma)-f_-(\Gamma)}{f_+(\Gamma)(d+|\Delta|)^{-1}+f_-(\Gamma)(d-|\Delta|)^{-1}
+C e^{-\Gamma |\Delta|}}\;,
\label{DEu_sol}
\end{equation}
where $f_{\pm}(\Gamma)=(y_0 \pm d) \exp[\pm(\Gamma-\Gamma_0)d]$ and the value of the
constant, $C$, follows from the boundary condition at $\Gamma=\Gamma_0$. 

Having the solution at hand  first, we check that
at the critical point, where $\Delta(\lambda)=0$ and thus $d^2=D^2=c^2$ we recover
the previous solution. Indeed, the constant in Eq.(\ref{DEu_sol})
at the critical point is given by $C=[(y_l^0)^2-d^2]^{1/2}$ and than combining Eq.(\ref{DEy_sol})
with Eq.(\ref{DEu_sol}) in the small $\Omega$ and $\lambda$ scaling limit we recover
the result in Eq.(\ref{p-pi-sol}).

In the Griffiths phases, $|\Delta(\lambda)|>0$, keeping in mind Eq.(\ref{lengths}) we have in the
small $\lambda$ limit:
\begin{equation}
d(\lambda)-|\Delta(\lambda)|=\frac{d(\lambda)^2-\Delta(\lambda)^2}{d(\lambda)+|\Delta(\lambda)|}
=A\lambda |\Delta(\lambda=0)|+O(\lambda^2)\;,
\label{A}
\end{equation}
so that the appropriate scaling combination in Eqs. (\ref{DEy_sol}) and (\ref{DEu_sol})
as $\Omega \to 0$ and $\lambda \to 0$ is $\lambda (\Omega_0/\Omega)^{2|\Delta|}=O(1)$.
The constant, $A$, in Eq.(\ref{A}) is related to the average cluster-size at $\Omega=\Omega_0$,
$\overline{l}_s(\Omega_0)$, so that finally we obtain along the line of semi-critical points:
\begin{equation} 
u_l=\frac{(y_l^0+|\Delta|)^2(\Omega_0/\Omega)^{2|\Delta|} \lambda/(2 |\Delta|) \overline{l}_s(\Omega_0)}
{(y_l^0+|\Delta|)^2(\Omega_0/\Omega)^{2|\Delta|} \lambda (1/4\Delta)^2 \overline{l}_s(\Omega_0)+1}\;.
\label{DEu_sol_sc}
\end{equation}
Now the joint distribution of the fields (couplings) and lengths can be obtained
through inverse Laplace transformation of Eqs. (\ref{lapl-tr}) using
Eqs.(\ref{P-lamb-sol}) and (\ref{R-lamb-sol}) and the solutions in Eqs. 
(\ref{p-r}), (\ref{pr-pirho}), (\ref{DEy_sol}) and (\ref{DEu_sol_sc}).

\end{multicols}
\widetext


\begin{references}
\bibitem{remark}
	The first equation in Eq.(\ref{magn-d}) does not give the complete $\Delta$
	dependence of the average magnetization, since the constant in this formula
	depends on the non-universal parameter, $\overline{y}$, which could also
	be of $\Delta$ dependent.
\end{references}
\end{document}